\documentclass{ws-mpla}

\begin{document}

\markboth{Merab Gogberashvili} {Gravitational Field of Spherical
Branes}

\catchline{}{}{}{}{}

\title{GRAVITATIONAL FIELD OF SPHERICAL BRANES
}

\author{\footnotesize MERAB GOGBERASHVILI}

\address{Andronikashvili Institute of Physics, 6 Tamarashvili Street, Tbilisi 0177, Georgia \\
and \\
Javakhishvili State University,
3 Chavchavadze Avenue, Tbilisi 0128, Georgia \\
e-mail: gogber@gmail.com}

\maketitle

\pub{Received 24 July 2008}{}

\begin{abstract}
The warped solution of Einstein's equations corresponding to the
spherical brane in five-dimensional AdS is considered. This metric
represents interiors of black holes on both sides of the brane and
can provide gravitational trapping of physical fields on the
shell. It is found the analytic form of the coordinate
transformations from the Schwartschild to co-moving frame that
exists only in five dimensions. It is shown that in the static
coordinates active gravitational mass of the spherical brane, in
agreement with Tolman's formula, is negative, i.e. such objects
are gravitationally repulsive.

\end{abstract}

\ccode{PACS Nos.: 04.50.+h, 04.20.Jb, 11.27.+d}

\vskip 1cm


\noindent 
Motivated by string/M theory,\cite{Hor-Wit} the AdS/CFT
correspondence\cite{Mal} and the hierarchy problem of particle
physics,\cite{ADD-1,ADD-2,AADD} in recent years there was a lot of
activity in brane-world models. In simplest brane-world scenario
our universe is considered as a boundary of a 5D AdS bulk with
induced Minkowski metric on the 
brane\cite{Gog-1,Gog-2,Ran-Sun-1,Ran-Sun-2}. There also exist various
warped co-dimension-$1$ models with curved brane in AdS, e.g. the
spatially flat cosmological brane-worlds,\cite{Shell-1,Shell-2,Shell-3,Shell-4,Shell-5,Shell-6} or the
spherical static ones.\cite{AdS-Black-1,AdS-Black-2,AdS-Black-3,AdS-Black-4,AdS-Black-5,Asymmetry-1,Asymmetry-2,Asymmetry-3}

In thin-wall approximation the brane is considered as a
hypersurface with $\delta$-like singularity in its
energy-momentum density
\begin{equation}
T^A_B = \delta \left( \frac z\epsilon \right)~ diag(T^0_0, T_1^1,
T_2^2,T_3^3, 0)~,
\end{equation}
where $\epsilon$ is the length scale associated with the brane.
Pressure towards the extra space-like dimension $z$ is assumed to
be zero. Capital Latin indices refer to 5D with the signature
$(+ - - - -)$. It is usually assumed that flat branes, which may
also be considered as a domain walls, are described by the state
equation\cite{Vil-She}
\begin{equation}\label{state}
T^0_0 = - T_1^1 = - T_2^2 = - T_3^3 = \frac \sigma \epsilon =
const~,
\end{equation}
where $\sigma >0 $ is the wall tension.

In the definition of active gravitational mass by Tolman's formula
\begin{equation}\label{m-active}
m_{active} =\int\sqrt{g}~d^4V(T_0^0 - T_1^1 - T_2^2 - T_3^3)  = -2
\sigma \int \sqrt{-g^{(4)}}~d^3V ~,
\end{equation}
where $\sqrt{g},~ d^4V$ are 5D and $\sqrt{-g^{(4)}},~ d^3V$ are
4D determinants and volume elements respectively, pressure terms
act as a repulsive source of gravity and flat branes have negative
gravitational mass.\cite{Vil-She} Note that other topological
objects also have peculiar gravitational properties. It was found
that cosmic strings do not produce gravitational force on the
surrounding matter locally, while global monopoles and global
strings are repulsive.\cite{Vil-She}

Einstein's equations in five dimensions,
\begin{equation} \label{Einstein}
R_{AB} - \frac 12 ~g_{AB} R = - \Lambda g_{AB} + \frac 1{M^3}~
T_{AB}~,
\end{equation}
where $M$ is the fundamental scale and $\Lambda > 0$ is bulk
cosmological constant (corresponding to AdS),  in the outer
regions ($T_{AB} = 0$) to a flat brane have the well known
solution\cite{Gog-1,Gog-2,Ran-Sun-1,Ran-Sun-2}
\begin{equation} \label{GRS}
ds^2 = e^{-|z|/\epsilon} dl^2 - dz^2~.
\end{equation}
The width of the brane $\epsilon$ relates to the brane tension,
fundamental scale and bulk cosmological constant by the fine-tune
conditions
\begin{equation} \label{fine-tune}
\epsilon^2 = \frac 6 \Lambda = \frac{M^2}{6\sigma}~.
\end{equation}
In this so-called warped setup (\ref{GRS}) the curvature is in the
bulk but the brane is flat, i.e. the induced metric on the brane
$dl^2$ is the ordinary 4D-Minkowski.

Curved branes in warped background are of much interest from the
cosmological and gravitational point of view. In this paper we
consider gravitational properties of a spherical 3D-brane
embedded in a 5D AdS. It will be natural to think that this
object, similarly to a flat brane, manifests gravitational
repulsion, since it is assumed that flat and spherical domain
walls obey the same state equations (\ref{state}) (e.g. see
Refs. 25 and 26). On the other hand, according to Birkhoff's
theorem, the empty space surrounding any spherical body is
described by the Schwarzschild metric, where the parameter
corresponding to the mass of gravitating body is usually assumed
to be positive. The disagreements in gravitational properties of
planar and spherical domain walls were explained by instability of
the latter,\cite{Ips-Sik} or by the existence of a positive energy
source stabilizing the bubble.\cite{Lop} However, there still
remain various paradoxes\cite{Ips-Sik,Gro,BGG} that can be solved
only if bubbles with the state equation (\ref{state}) are
repulsive.

It is also possible that domain walls are not described by the
state equation (\ref{state}), since there exist several problems
within models with large pressure (see recent papers\cite{pressure-1,pressure-2}). If nevertheless the state equation
for a spherical object has the form (\ref{state}), it is still
possible to construct a model where repulsive bubbles are
admissible. For example, one can consider reversion of time on the
shell (analogously to what happens between the upper and Cauchy
radii in Reissner-Nordstr\"{o}m metric),\cite{Bar-Gog-1} or use
non-Minkowskian asymptotic.\cite{Bar-Gog-2}

The setup we consider here, 3D spherical brane in 5D AdS, is
similar to static AdS black hole models\cite{AdS-Black-1,AdS-Black-2,AdS-Black-3,AdS-Black-4,AdS-Black-5}.
Unlike to these models, we do not restrict ourselves with the
standard AdS-Schwarzschild solution (with the positive mass
parameter), since in this case spherical domain walls do not
exhibit gravitational trapping without extra sources.\cite{Rah-Cha-1,Rah-Cha-2} Instead we consider the metric {\it
ansatz}
\begin{equation} \label{Ansatz}
ds^2 = B^2 (a) ~dt^2 - da^2 - A^2 (a)~ d\Omega^2_3~,
\end{equation}
in the similar to (\ref{GRS}) holographic-like frame (i.e. in
Gaussian coordinates) and later check to what kind of geometry in
co-moving coordinates it corresponds. In (\ref{Ansatz}) the metric
components $A(a)$ and $B(a)$  are functions of the 5D radial
(holographic) coordinate $a$ only, and
\begin{equation}
d\Omega^2_3 = d \kappa + \sin^2\kappa~ d\theta^2 + \sin^2\kappa ~
\sin^2\theta~d\phi^2
\end{equation}
is the differential solid angle of the unit three-sphere with the
three spherical angles $\kappa, \theta$ and $\phi$.

In the Gaussian coordinates, the spherical brane is located at $a =
R$. It is known that domain wall's metric with planar, or
spherical symmetries is not in general static, but admits a de
Sitter-like expansion.\cite{Accel-wall-1,Accel-wall-2,Accel-wall-3,Accel-wall-4,Accel-wall-5,Accel-wall-6}
However, the problem we address in this paper is to find warped
solution and active gravitational mass for spherical branes (as
was mentioned above in a static coordinates they are expected to be
gravitationally repulsive). So we consider static coordinates,
where the radius of the bubble $R$ is assumed to be constant.

For the {\it ansatz} (\ref{Ansatz}) Einstein's field equations
reduce to
\begin{eqnarray} \label{System}
\frac{A^{\prime 2}}{A^2} + \frac{A^{\prime\prime}}{A} -
\frac{1}{A^2} &=& \frac \Lambda 3~,
\nonumber \\
\frac{A^{\prime 2}}{A^2} +
\frac{A^{\prime}}{A}\frac{B^{\prime}}{B} -
\frac{1}{A^2} &=& \frac \Lambda 3~, \\
\frac{A^{\prime 2}}{A^2} + 2
\frac{A^{\prime}}{A}\frac{B^{\prime}}{B} + 2
\frac{A^{\prime\prime}}{A} + \frac{B^{\prime\prime}}{B}-
\frac{1}{A^2} &=& \Lambda~,\nonumber
\end{eqnarray}
where primes denote derivatives with respect of $a$. From the
first two equations, it follows that
\begin{equation} \label{B}
B = C_1 A^\prime ~,
\end{equation}
where $C_1$ is the integration constant. Then for the second
unknown function we find
\begin{equation} \label{A}
A^2 = C_2 e^{- a/\varepsilon} + C_3 e^{a/\varepsilon} - 2
\varepsilon^2 ~,
\end{equation}
where $C_2$ and $C_3$ are additional integration constants. We
also introduced the parameter
\begin{equation} \label{varepsilon}
\varepsilon = \sqrt{\frac{3}{2\Lambda}}~,
\end{equation}
corresponding to the width of the brane.

Note that the solution represented by (\ref{B}) and (\ref{A}) is
not new, it is just a special case (corresponding to the static
brane) of cosmological brane-world solutions extensively discussed
in the literature (see, e.g., Refs. 12-14). Our choice of boundary 
conditions is new. In (\ref{B}) and (\ref{A})
we want to choose integration constants $C_1, C_2$ and $C_3$ in
order to receive the warped metric that on the surface of junction
(on the spherical shell with the radius $R$) will coincide with
the Minkowski metric. Recognizing the difficulty of handling thick
walls within relativity we, as many other authors, will use
thin-wall formalism.\cite{BKT,Isr} To obtain warped metric for
the shell, we need decreasing functions $A^2$ and $B^2$ obeying 
the boundary conditions
\begin{equation} \label{boundary}
A^2 (a = R) = R^2 ~, ~~~ B^2 (a = R) = 1~.
\end{equation}
In general, in the first expression we can multiply $R^2$ by some
constant that will correspond to a 3D shell with a solid angle
deficit.

The boundary conditions (\ref{boundary}) fix the values of
integration constants for the outer region to the shell
\begin{eqnarray}
C^+_1 &=& \frac{2\varepsilon R}{R^2 + 2 \varepsilon^2}~, \nonumber
\\
C^+_2 &=& (R^2 + 2 \varepsilon^2) e^{R/\varepsilon}~, \\
 C^+_3 &=& 0~. \nonumber
\end{eqnarray}
Thus the metric functions  outside the shell ($a>R$) are
\begin{eqnarray} \label{outer}
A_+^2 &=& (R^2 + 2 \varepsilon^2) e^{- (a-R)/\varepsilon}  - 2
\varepsilon^2 ~,\nonumber \\ B_+^2 &=& \frac{R^2}{A_+^2}e^{-
2(a-R)/\varepsilon}~.
\end{eqnarray}

Note that because of warped background the usual cancellation of
Newton's gravitational potential inside a spherical shell does not
occur and we can assume the $Z_2$-symmetry between the two sides
of the shell. Then integration constants inside can be chosen as
\begin{eqnarray}
C^-_1 &=& \frac{2\varepsilon R}{R^2 + 2 \varepsilon^2}~, \nonumber
\\
C^-_2 &=& 0 ~, \\
 C^-_3 &=& (R^2 + 2 \varepsilon^2)
e^{-R/\varepsilon}~, \nonumber
\end{eqnarray}
and the metric functions in the interior region ($a<R$) have a
form similar to (\ref{outer})
\begin{eqnarray} \label{inter}
A_-^2 &=& (R^2 + 2 \varepsilon^2) e^{(a-R)/\varepsilon}  - 2
\varepsilon^2 ~, \nonumber \\ B_-^2 &=& \frac{R^2}{A_-^2}e^{
2(a-R)/\varepsilon}~.
\end{eqnarray}
One can choose another solution inside the shell, e.g. 5D
Minkowski, or Minkowski-AdS. We don't consider these
possibilities, since in this paper we want to show existence of
warping for a spherical brane that approximates the flat one
(\ref{GRS}) when the radius of the shell is large.

Note that our metric (\ref{Ansatz}), similarly to the case of
asymmetrically warped space-times models,\cite{Asymmetry-1,Asymmetry-2,Asymmetry-3} contains two different
warp factors $A^2_\pm$ and $B^2_\pm$, one associated with time and
another with 3D-space. This violates Lorentz invariance\cite{LV-1,LV-2} and in general light speed on the shell and in
the bulk will be different.\cite{Speed}

The metric functions $A_\pm(a)$ in (\ref{outer}), (\ref{inter})
become zero at some distance $d$. Thus space-time of the shell
with $Z_2$ symmetry has singularities (horizons) in both, interior
and exterior regions, which can be cancelled by the introduction
of some sources in the bulk. We can estimate the distance to these
singularities
\begin{equation}
d = \varepsilon ~\ln \left( \frac{R^2 + 2 \varepsilon^2}{2
\varepsilon^2}\right) \sim 100 ~\varepsilon
\end{equation}
if we use, for example, the values of the radius of curvature of
our universe $R \sim 10^{28} ~cm$ and experimentally acceptable
width of the brane $\varepsilon \sim 10^{-3}~cm$.

The second metric functions $B^2_\pm$ ('potential') in
(\ref{outer}), (\ref{inter}) exhibits slightly different behavior.
It decreases from the brane location to it's minimal value
\begin{equation}
B^2_{min}(a - R = d) = \frac{8 \varepsilon^2 R^2}{(R^2 + 2
\varepsilon^2)^2}~,
\end{equation}
(that is a very small number) at the distance $d-\varepsilon\ln2$.
Then it starts to grow up and approaches infinity at $d$.

Using the expressions (\ref{outer}) and (\ref{inter}) our {\it
ansatz} (\ref{Ansatz}) can be rewritten in a form similar to
(\ref{GRS}) with the common warp factor for 4D part of the
metric
\begin{equation} \label{warp-ds}
ds^2_\pm =  \left[ \frac{R^2 + 2 \varepsilon^2}{a^2}
e^{\mp(a-R)/\varepsilon} - \frac{2 \varepsilon^2}{a^2}
\right]\left[ U^2_\pm(a) dt^2 - a^2 d\Omega^2_3 \right] - da^2~,
\end{equation}
where the values of the functions
\begin{equation} \label{U}
U(a)_\pm = \frac{Ra  e^{\mp(a-R)/\varepsilon}}{(R^2 + 2
\varepsilon^2) e^{\mp(a-R)/\varepsilon}- 2 \varepsilon^2}~,
\end{equation}
close to the brane ($a \approx R$) are approximately equal to $1$.
As it was expected, in the large-shell limit, $a \sim R \gg
\varepsilon$, singularities of the metric (\ref{warp-ds}) are
shifted to the infinity and (\ref{warp-ds}) transforms to the
metric of a flat brane (\ref{GRS}).

The determinant in our {\it ansatz} (\ref{Ansatz}) is given by
\begin{equation} \label{determinant}
\sqrt{g} = \sqrt{-g^{(4)}} \frac{A^3(a)B(a)}{ R^3}~,
\end{equation}
where $ \sqrt{-g^{(4)}}$ is the 4D determinant on the shell.
Localization of the 4D spin-$2$ graviton on the spherical brane
requires the integral of the gravitation part of the action,
\begin{equation}
S = \int d^5 x\sqrt{g} ~\frac{M^3}{2}R~,
\end{equation}
over $a$ to be convergent. If we neglect the effects of $U(a)_\pm$
our metric (\ref{warp-ds}) will have the conformal structure
similar to (\ref{GRS}) and only the integral of the determinant
(\ref{determinant}) will be nontrivial. In this case for the
effective Planck's scale on the shell we find
\begin{equation} \label{Planck}
m^2_{pl} \approx \frac{M^3}{R^3} \int da ~A^3 B =
\frac{4\varepsilon M^3}{R^2(R^2 + 2 \varepsilon^2)} \int_0^R dA~
A^3 = \frac{\varepsilon R^2 M^3}{(R^2 + 2 \varepsilon^2)} \approx
\varepsilon M^3,
\end{equation}
which is the standard expression for the large extra dimensional
models.\cite{ADD-1,ADD-2,AADD} In the case of shell with a
deficit angle in (\ref{Planck}) a new parameter will appear and
one will be able to explain large hierarchy even in 5D.

Now we can translate our solutions (\ref{outer}), (\ref{inter})
into static co-moving coordinates  by introducing the new 'radial'
coordinates
\begin{equation} \label{rho}
r_\pm = A_\pm~,
\end{equation}
which are connected to the proper radius $a$ by
\begin{equation} \label{r}
a = R + \varepsilon \ln \left( \frac{R^2 + 2\varepsilon^2}{r_+^2 +
2\varepsilon^2}\right) = R + \varepsilon \ln \left( \frac{r_-^2 +
2\varepsilon^2}{R^2 + 2\varepsilon^2}\right)~.
\end{equation}
Then (\ref{Ansatz}) transforms to Schwarzschild's coordinates
\begin{equation} \label{co-moving}
ds^2 = D(r_\pm) dt^2 - \frac{dr^2_\pm}{D(r_\pm)} - r^2_\pm
d\Omega_3^2~,
\end{equation}
where the only independent metric function has the form
\begin{equation} \label{D}
D(r_\pm) = 1 + \frac{\varepsilon^2}{r^2_\pm} + \frac{r^2_\pm}{4
\varepsilon^2} ~.
\end{equation}
We see that horizons of the metric (\ref{warp-ds}) in the new
coordinates correspond to the central singularities of the
Schwarzschild metrics at $r_\pm = 0$. So spherical brane in our
setup can be imagined as the common boundary (event horizon) of
two black holes situated at different sides of the shell.

Recalling the definition  of the brane width (\ref{varepsilon}), we
recognize that the last term in (\ref{D}),
\begin{equation}
\frac{r^2_\pm}{4 \varepsilon^2} = \frac{\Lambda_\pm}{6}r^2_\pm~,
\end{equation}
is the standard term with the cosmological constant in 5D
AdS-Schwarzschild metric.

Note that the value and the sign of the second term
should represent the gravitational potential in five dimensions.
Because of the positive value of the potential, we conclude that warped
spherical brane in AdS should have negative active gravitational
mass equal to
\begin{equation} \label{m-brane}
m_{brane} = - \frac{3 M^3}{2 \Lambda}~.
\end{equation}
This means that static spherical brane with warped geometry should
be gravitationally repulsive in agreement with Tolman's formula
(\ref{m-active}). Using the formulae (\ref{m-active}) and
(\ref{determinant}), and the expression for the volume of
3D-sphere ($V^3 = 2\pi^2 R^3$), from (\ref{m-brane}) one can
find that the brane tension is related to the shell radius,
fundamental scale and bulk cosmological constant as
\begin{equation}
\sigma = \frac{3 M^3}{8\pi^2 R^3 \Lambda}~.
\end{equation}
This is analogue to the fine-tuning condition (\ref{fine-tune}) in
the flat brane model (\ref{GRS}).

In this paper the warped solutions (\ref{outer}), (\ref{inter}) of
Einstein's equations corresponding to the spherical brane in 5D
AdS is considered. These solutions in Schwarzschild's coordinates
represent interiors of black holes placed at both sides of the
brane. So in realistic models one needs some sources in the bulk
to cancel these singularities and also to stabilize the shell. It
is shown that spherical branes in static coordinates are
gravitationally repulsive in agreement with the estimations of
Tolman's formula (\ref{m-active}). Existence of warping in our
model means that matter fields can be trapped gravitationally by
spherical branes, similarly to the case of flat ones. Note that
our analytic solutions, and simple transformation to co-moving
coordinates (\ref{r}), takes place only in five dimensions, where
the orders of the second and third terms in the Schwarzschild
metric function (\ref{D}) (corresponding to the gravitational
potential and the cosmological constant respectively) coincide.


\end{document}